\documentclass[preprintnumbers,superscriptaddress,a4paper]{revtex4}
\usepackage{amssymb}
\usepackage{amsfonts}
\usepackage{amsmath}
\usepackage{graphicx}
\usepackage{dcolumn}
\usepackage{bm}

\begin{document}

\title{Neuronal avalanches of a self-organized neural network with active-neuron-dominant structure}

\author{Xiumin Li}
\email[]{email: freexmin@gmail.com}
\affiliation{College of Automation, Chongqing University, Chongqing 400044, China}
\affiliation{Department of Electronic and Information Engineering, Hong Kong Polytechnic University, Hung Hom, Kowloon, Hong Kong}

\author{Michael Small}
\affiliation{School of Mathematics and Statistics, University of Western Australia, Crawley, WA 6009, Australia}
\affiliation{Department of Electronic and Information Engineering, Hong Kong Polytechnic University, Hung Hom, Kowloon, Hong Kong}

\begin{abstract}

Neuronal avalanche is a spontaneous neuronal activity which obeys a power-law distribution of population event sizes with an exponent of $-3/2$. It has been observed in the superficial layers of cortex both \emph{in vivo} and \emph{in vitro}. In this paper we analyze the information transmission of a novel self-organized neural network with active-neuron-dominant structure. Neuronal avalanches can be observed in this network with appropriate input intensity. We find that the process of network learning via spike-timing dependent plasticity dramatically increases the complexity of network structure, which is finally self-organized to be active-neuron-dominant connectivity. Both the entropy of activity patterns and the complexity of their resulting post-synaptic inputs are maximized when the network dynamics are propagated as neuronal avalanches. This emergent topology is beneficial for information transmission with high efficiency and also could be responsible for the large information capacity of this network compared with alternative archetypal networks with different neural connectivity.

\end{abstract}

\maketitle

\textbf{Recent neurophysiological experiments report that scale-invariant dynamics known as neuronal avalanches can be broadly observed in spontaneous cortical activities. These dynamics exhibit a mixture of ordered and disordered patterns and are believed to be highly efficient for information processing in the neocortex. In this paper we investigate neuronal avalanches in a novel active-neuron-dominant neural network which is self-organized by spike-timing-dependent plasticity --- a model of neural development which is also strongly supported by direct physiological evidence. The refinement process of network learning selectively strengthens or weakens synaptic connections based on the intrinsic excitability of individual neurons. We find that neuronal avalanches can be observed in this network with medium external excitation. Moreover, activity entropy measured from population dynamics is maximized when neuronal avalanches exist. Meanwhile, the active-neuron-dominant network connectivity dramatically enhances the activity entropy of neuronal population and displays large complexity of network structure compared with alternative archetypal networks with different neural connectivity. This result may have important implications on understanding the potential functions of network learning for improving the efficiency of information processing.}

\section{Introduction}

It has been widely believed that brain dynamics are collective processes involving synaptic integrations from thousands of neurons in cortical networks. There is evidence showing that healthy spontaneous brain dynamics originating from collective processes are not composed of either completely random activity patterns or periodic oscillations \cite{buzs¨¢ki2006rhythms,chialvo2010emergent}. Recent studies by Beggs and Plenz reported a type of spontaneous activity with critical dynamics, where spatiotemporal patterns are distributed in sizes according to a power law with a slope of $-1.5$ \cite{beggs2003neuronal,beggs2004neuronal,petermann2009spontaneous}. These experimental results exactly agree with the mean-field theoretical analysis for self-organized criticality (SOC), which is a critical branching process observed in many systems \cite{bak1987self,zapperi1995self}. This critical dynamic represents spatially irregular patterns of propagated medium synchronization in neural circuits. It has been found in superficial layers of cortex both \emph{in vivo} and \emph{in vitro} \cite{beggs2003neuronal,plenz2007organizing,gireesh2008neuronal,pasquale2008self,petermann2009spontaneous,tetzlaff2010self}. Neuronal avalanches play a substantial role in cortical circuits, especially for information transmission. It has been shown that there is a close relationship between SOC and synchronization in neural networks \cite{bottani1995pulse}. Information capacity and transmission, as well as dynamic range, are maximized in cortical networks with neuronal avalanches \cite{shew2009neuronal,shew2011information}.

Computational models with respect to neuronal avalanches have been investigated in recent years. The small-world topology was identified from network reconstruction of experimental data with neuronal avalanches \cite{pajevic2009efficient}. A neuronal network model with scale-free topology was developed to investigate the generation of avalanches distribution \cite{pasquale2008self}. Besides these networks with predefined structures, self-organized neural networks with synaptic plasticity can also achieve critical neuronal dynamics \cite{bak2001adaptive,shin2006self,teramae2007local,de2006self,shin2006self,levina2007dynamical,
abbott2007simple,meisel2009adaptive,millman2010self}. However, few studies consider the relationship or the interaction between the development of network structure and the dynamics of neuronal population. In our recent papers \cite{li2009self,li2010enhancement} we proposed a novel neural network refined from spike-timing-dependent plasticity (STDP). Due to the existence of heterogeneity in the excitability degrees of neurons, the network finally evolves into a sparse and active-neuron-dominant structure. That is, strong connections are mainly distributed to the outward links of a few highly active neurons. Such synapse distribution renders these active cells a powerful drive to trigger the other neurons firing synchronously, thus spreading the excitation of the whole network activity. The refinement of synaptic connectivity encodes well the heterogeneity of intrinsic dynamic and significantly promotes network synchronization \cite{li2009self}. A recent experimental study \cite{yassin2010embedded} have found that a small population of highly active neurons may dominate the firing in neocortical networks, suggesting the existence of active-neuron-dominant connectivity in the neocortex.

In this paper neuronal avalanches have been observed in a self-organized neural network with active-neuron-dominant structure by changing the applied external excitation. This finding is consistent with the experimental results that cortex operates in critical regime with appropriate local excitability \cite{plenz2007organizing}. Both the entropy of activity patterns and the complexity of their resulting post-synaptic inputs are maximized when the network dynamics are propagated as neuronal avalanches. We find that the process of activity-dependent synaptic plasticity dramatically increases the complexity of network structure, which is self-organized as active-neuron-dominant connectivity after the learning process. This emergent network has high complexity of network topology which benefits for enhancing information transmission of neural circuits.

\section{self-organized neural network with synaptic plasticity}

Here the self-organization of the neuronal network through the STDP learning is briefly introduced. The network used in this paper is composed of $100$ synaptically connected regular spiking neurons which are modeled by the two-variable IF model of Izhikevich \cite{izhikevich2003simple}. It is described by
\begin{equation}\label{neuron}
 \begin{array}{lll}
 \dot{v_{i}}=0.04v_{i}^2+5v_{i}+140-u_{i}+I+I_{i}^{syn} \\
 \dot{u_{i}}=a(bv_{i}-u_{i})+D\xi_{i}
 \end{array}
\end{equation}
\begin{equation}
 \mbox{if $v_{i}>30$ mV, then }\left\{
 \begin{array}{lll}
 v_{i} \leftarrow c\\
 u_{i} \leftarrow u_{i}+d
 \end{array}
 \right.
\end{equation}
where $i=1,2,...,N$, $v_{i}$ represents the membrane potential and
$u_{i}$ is a membrane recovery variable. The parameters $a,b,c,d$
are dimensionless. The variable $\xi_{i}$ is the independent
Gaussian noise with zero mean and intensity $D$ that represents the
noisy background. $I$ stands for the externally applied current, and
$I_{i}^{syn}$ is the total synaptic current through neuron $i$ and
is governed by the dynamics of the synaptic variable $s_j$:
\begin{equation}
 \begin{array}{ll}
 I_{i}^{syn}=-\sum_{1(j\neq i)}^N g_{ji}s_{j}(v_{i}-v_{syn})\\
 \dot{s_{j}}=\alpha(v_{j})(1-s_{j})-s_{j}/\tau \\
 \alpha(v_{j})=\alpha_{0}/(1+e^{-v_{j}/v_{shp}})
 \end{array}
\end{equation}
here the synaptic recovery function $\alpha(v_{j})$ can be taken as
the Heaviside function. When the presynaptic cell is in the silent
state $v_{j}<0$, $s_j$ can be reduced to $\dot{s_{j}}=-s_{j}/\tau$;
otherwise $s_j$ jumps quickly to $1$ and acts on the postsynaptic
cells. The synaptic conductance $g_{ji}$ from the $j_{th}$ neuron to
the $i_{th}$ neuron will be updated through the STDP learning rule. Here the excitatory
synaptic reversal potential $v_{syn}$ is set to be $0$. The degree of neuron's excitability is governed by the parameter $b$. Neurons with larger $b$ are prone to exhibit larger excitability and fire with a higher frequency than others. In order to establish a heterogenous network, $b_{i}$ is uniformly distributed in $[0.12,0.2]$.

\begin{figure}
 \begin{center}
 \includegraphics[height=0.5\textwidth]{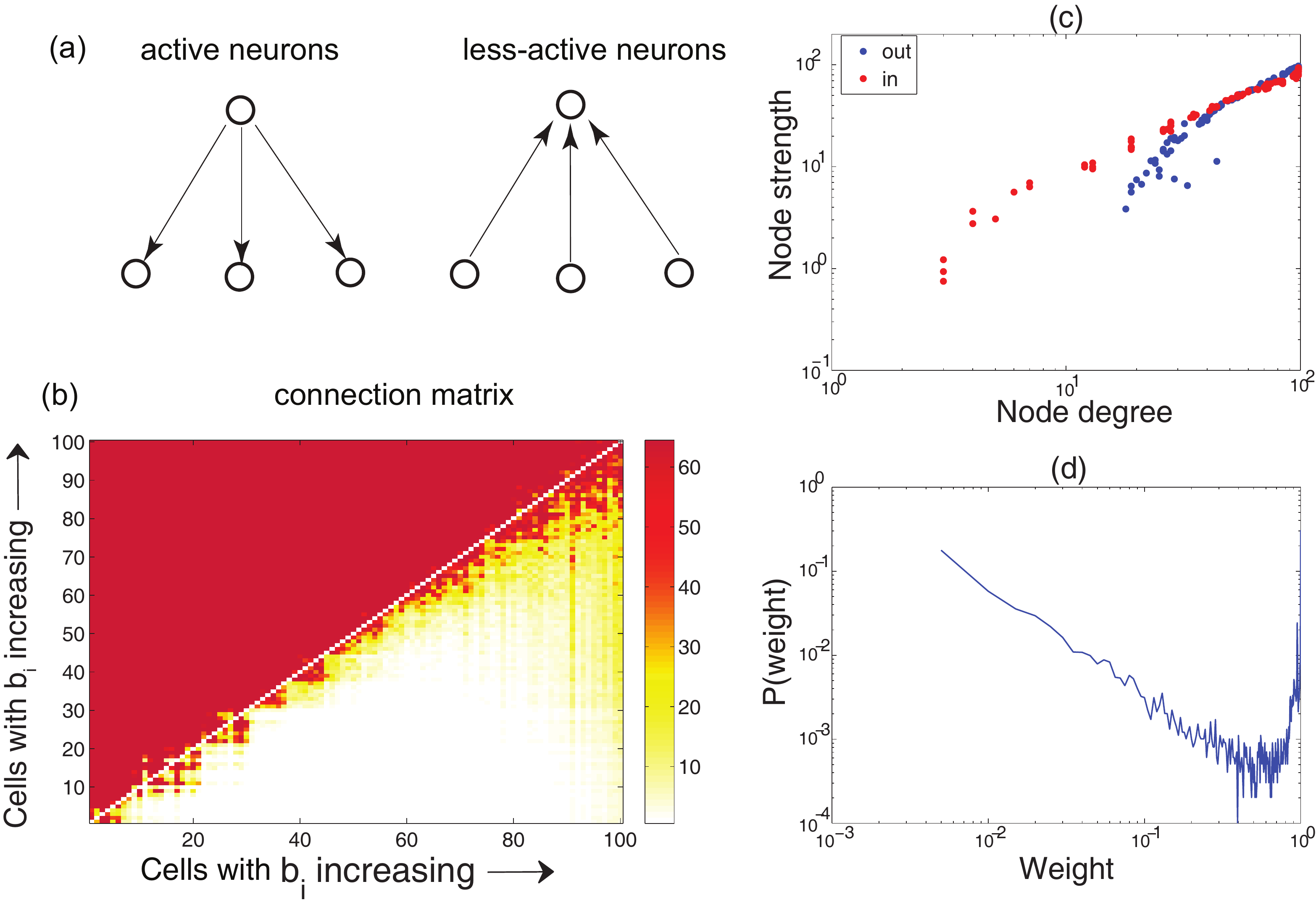}\\
 \caption{Self-organized neural network with active-neuron-dominant structure. (a) Left: Strong outward synaptic connections of active neurons with high excitability. Right: Strong inward synaptic connections of less-active neurons with low excitability. Arrows represent the directions of  synaptic connectivity. (b) Image of the normalized matrix of synaptic weights. Both x axis and y axis represent the index of neurons which is sorted with increasing value of the parameter $b$ (i.e. increasing excitability degree). (c) Node strength which is the summation of all inward or outward synaptic weights for each node is proportional to the node degree. (d) Probability distribution of synaptic weights is almost power law except for some extremely large synaptic weights.}\label{fig1}
 \end{center}
\end{figure}

In our simulation synapses between neighboring neurons are updated
by the STDP modification function $F$, which selectively strengthens
the pre-to-post synapses with relatively shorter latencies or
stronger mutual correlations, while weakening the remaining synapses
\cite{song2000competitive}. The synaptic
conductance is updated by
\begin{equation}\label{stdp}
 \begin{array}{ll}
 \Delta g_{ij}=g_{ij}F(\Delta t)\\
 F(\Delta t)=\left\{
 \begin{array}{ll}
 A_{+}\exp(-\Delta t/\tau_{+})~~\rm{if}~\Delta t>0\\
 -A_{-}\exp(\Delta t/\tau_{-})~~\rm{if}~\Delta t<0
 \end{array}
 \right.
 \end{array}
\end{equation} where $\Delta t=t_{j}-t_{i}$, $t_i$ and $t_j$ are the spike time of the presynaptic and postsynaptic cell respectively. $F(\Delta t)=0$ if $\Delta t=0$.
$\tau_{+}$ and $\tau_{-}$ determine the temporal window for synaptic
modification. The parameters $A_{+}$ and $A_{-}$ determine the
maximum amount of synaptic modification. Here, we set $\tau_{-}=\tau_{+}=20$,
$A_{+}=0.05$ and $A_{-}/A_{+}=1.05$ as used in \cite{song2000competitive}. The peak synaptic conductance is restricted to the range $[0,g_{max}]$, where $g_{max}=0.03$ is the limiting value. Other parameters used in this paper are $a=0.02, c=-65, d=8,
\alpha_{0}=3, \tau=2, V_{shp}=5, D=0.1$. The time step is $0.05$ ms.

In the initial state, each neuron is all-to-all bidirectionally connected with the same conductance of $g_{max}/2$. The whole network is subject to an external current ($I=6$) as a learning environment. After sufficient time (about $20s$) the connections evolve into a steady state and exhibit the locally active-neuron-dominant property as we have described in our previous work \cite{li2009self}. Most of the synapses are rewired to be either $0$ or $g_{max}$. Competition within this heterogenous network causes the active cells to have high out-degree synapses and low in-degree synapses, while the inactive ones are just the opposite (Fig.~\ref{fig1} (a) (b)). During the STDP learning process, the synaptic strengths of the network are renewed by increasing the influence of active cells over the others and the dependence of inactive cells on the active cells. Note that there is a nearly linear relationship between the node degree and node strength (Fig.~\ref{fig1} (c)), indicating the connections from active neurons to less-active neurons have both high degree and strong weight. The probability distribution of synaptic weights is almost power law except for some extremely large synaptic weights, suggesting the presence of hub-nodes with high capability of synaptic transmission (Fig.~\ref{fig1} (d)). In this way the internal dynamics of different neurons is encoded into the topology of the emergent network and therefore the communication between active neurons and inactive neurons is improved \cite{li2009self}. Since the network connectivity finally reaches stationary after STDP learning, this self-organized neural network is fixed without any update process and represented as STDP network in the following simulations (except for Fig.~\ref{fig8}).

\section{Neuronal avalanches of the self-organized neural network}

\begin{figure}
 \begin{center}
 \includegraphics[height=0.9\textwidth]{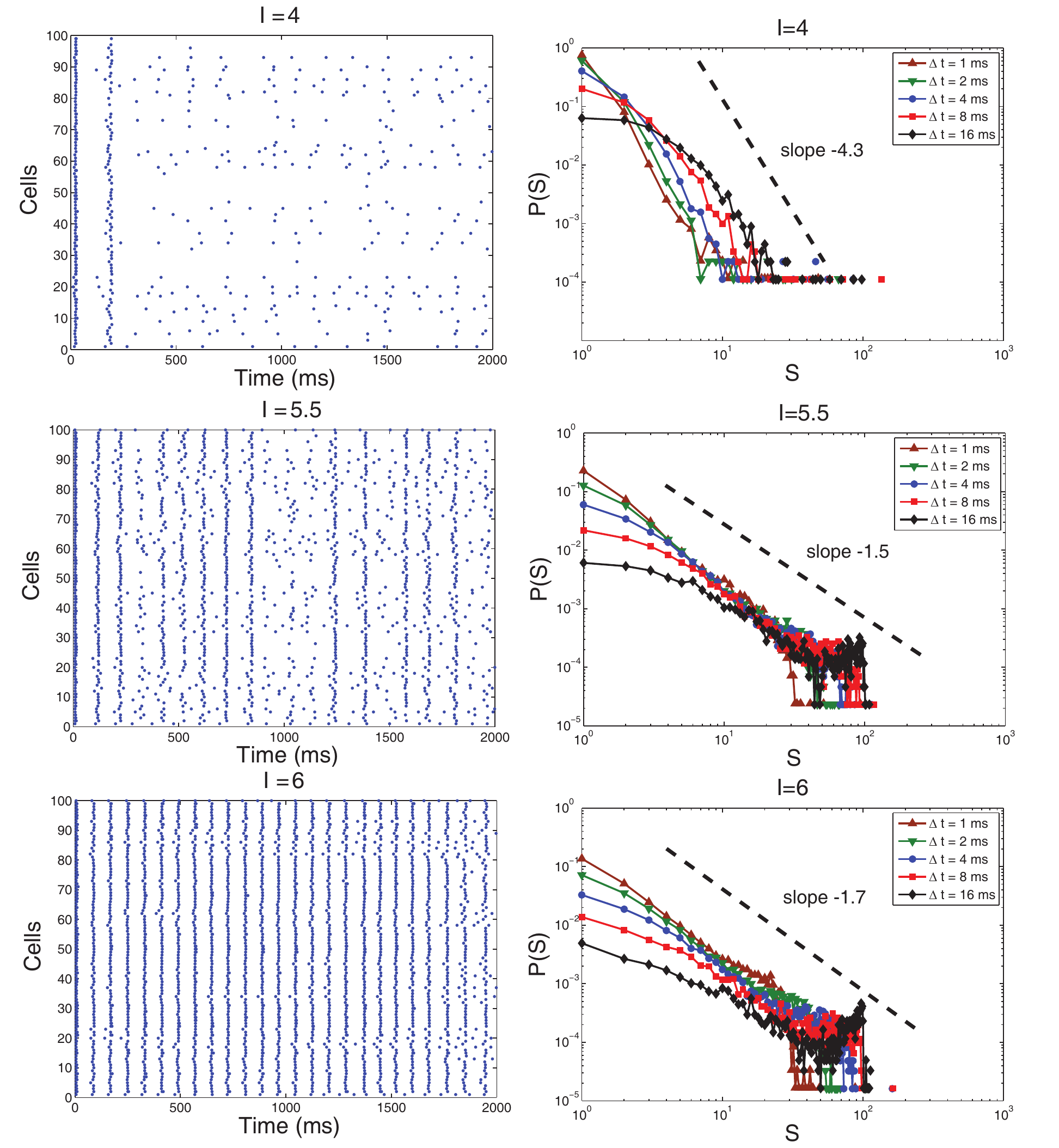}
 \caption{Raster of neuronal activities (left panel) and corresponding avalanche size distributions (right panel) for subcritical (top), critical (middle) and supercritical (bottom) dynamical state with different bin width $\Delta t$. The slope $\alpha$ for three cases when $\Delta t=4$~ms is calculated as the slope of linear function fitted from the nearly linear part of the distribution (with avalanche size ranged from about $2$ to $10$ for $I=4$, $4$ to $20$ for $I=5.5$ and $3$ to $30$ for $I=6$). }\label{fig2}
 \end{center}
\end{figure}

\begin{figure}
 \begin{center}
 \includegraphics[height=0.4\textwidth]{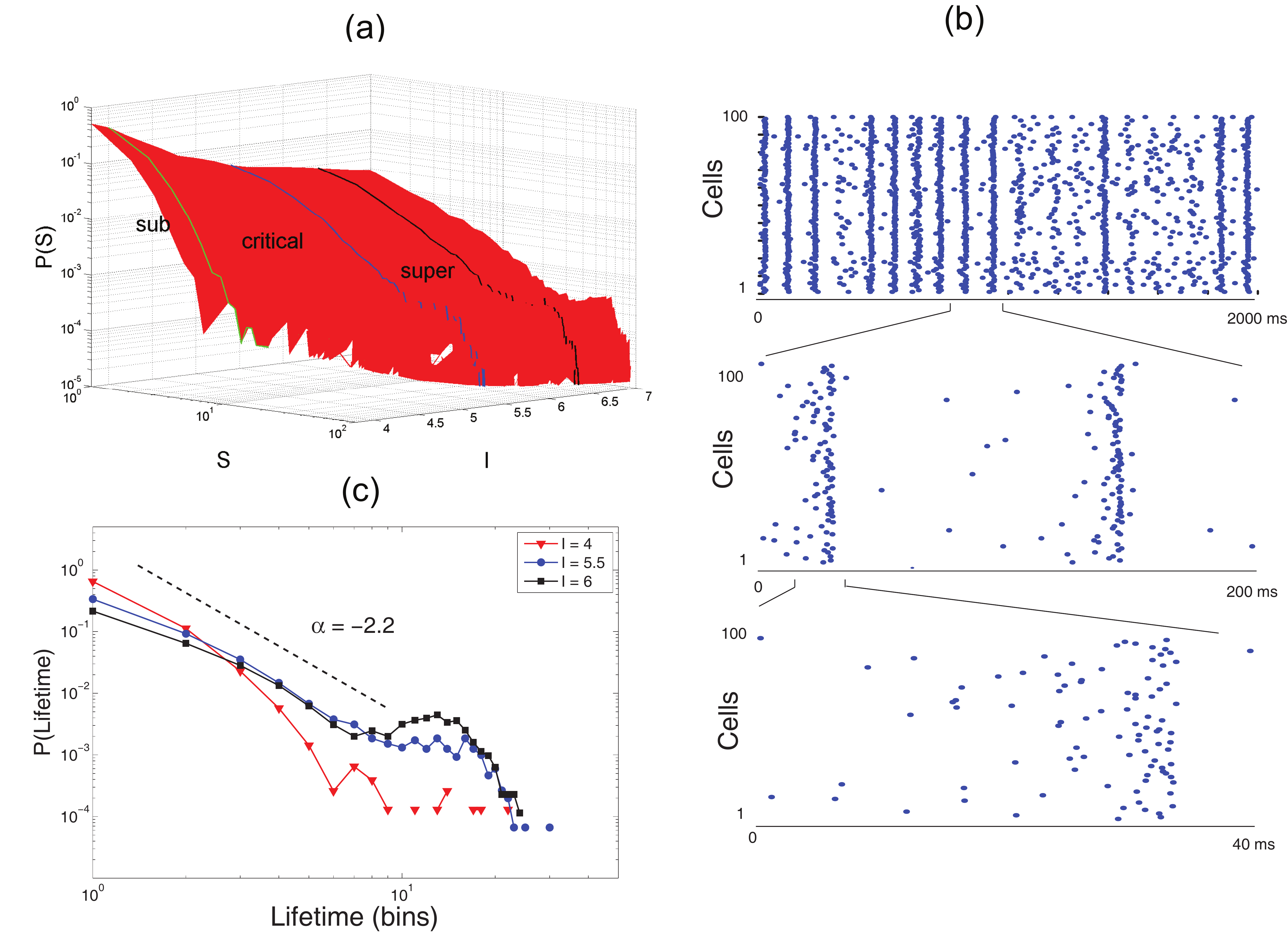}
 \caption{Neuronal avalanches of the STDP network. (a) Size distributions for avalanche sizes ranging from the subcritical state to supercritical state of the network activity by increasing the stimulus current ($I$). Avalanche size ($S$) is the number of neurons which are activated during each avalanche. Time bin width $\Delta t$ is $4$~ms. (b) Raster of neuronal activity with different spatiotemporal scales. (c) Avalanche lifetime distribution computed as the number of time bins in each avalanche. Fitting slope of the distribution for $I=5.5$ is about $-2.2$.}\label{fig3}
 \end{center}
\end{figure}

The left panel of Fig.~\ref{fig2} shows the network activity with different levels of external excitation. The average firing frequency and synchronization of network activity are improved by increasing the external current, which can be achieved by increasing the concentration of dopamine or the ratio of excitatory to inhibitory synaptic inputs in the cortex \cite{plenz2007organizing}. Due to the existence of synaptic connectivity the spatiotemporal patterns of these synchronized network activities are propagated through cascades that initiated from several excited neurons. A sequence of consecutive active frames that is preceded by a blank frame and ended by a blank frame is called an avalanche, where frame is the spatial pattern of active neurons during one time bin $\Delta t$ \cite{beggs2003neuronal}. Avalanche size ($S$) is the number of activated neurons during each avalanche. Probability distributions of avalanche sizes for corresponding network activities are shown in the right panel of Fig.~\ref{fig2}. The curves take on the form of a power law: $P(S)\propto S^{\alpha}$ when synchronized patterns are produced in the network activity (for cases $I=5.5$ and $I=6$ shown in Fig.~\ref{fig2}). Neuronal avalanche occurs when the network activity reaches medium synchronization with the slope $\alpha=-1.5$. Note that this scale-free property for the critical state of network synchrony is very robust for the change of time bin width $\Delta t$. In the following calculations, $\Delta t$ is set to be $4$~ms as in the studies \cite{beggs2003neuronal,beggs2004neuronal,pajevic2009efficient}. Fig.~\ref{fig3} (a) demonstrates that the propagation of dynamic patterns of this self-organized neural network can be transferred from the subcritical state, to critical and finally to the supercritical state by increasing the stimulus excitatory current ($I$). In the critical regime, large diversity of avalanche patterns can be found at many different scales (see Fig.~\ref{fig3} (b)). Moreover, the duration of an avalanche which is usually called avalanche lifetime and is expressed in number of time bins also follow a power law distribution when $I=5.5$ (Fig.~\ref{fig3} (c)). The exponent is about $-2.2$, very close to the value of $-2$ for lifetime distribution reported in \cite{beggs2003neuronal}. For the subcritical case, the exponential distribution indicates that only few neurons are activated during the excitation propagation, whereas in the supercritical state most of the network are activated synchronously, causing a long tail at the end of the distribution. These observations are consistent with the results obtained both experimentally and computationally \cite{beggs2003neuronal,pasquale2008self,pajevic2009efficient}. This critical state of information processing makes the network achieve medium synchronization neither to be incoherent spiking nor highly coherent activity, which indicates its high flexibility and sensitivity to external signals.


\section{Entropy of network activity and structure}

\begin{figure}
 \begin{center}
 \includegraphics[height=0.5\textwidth]{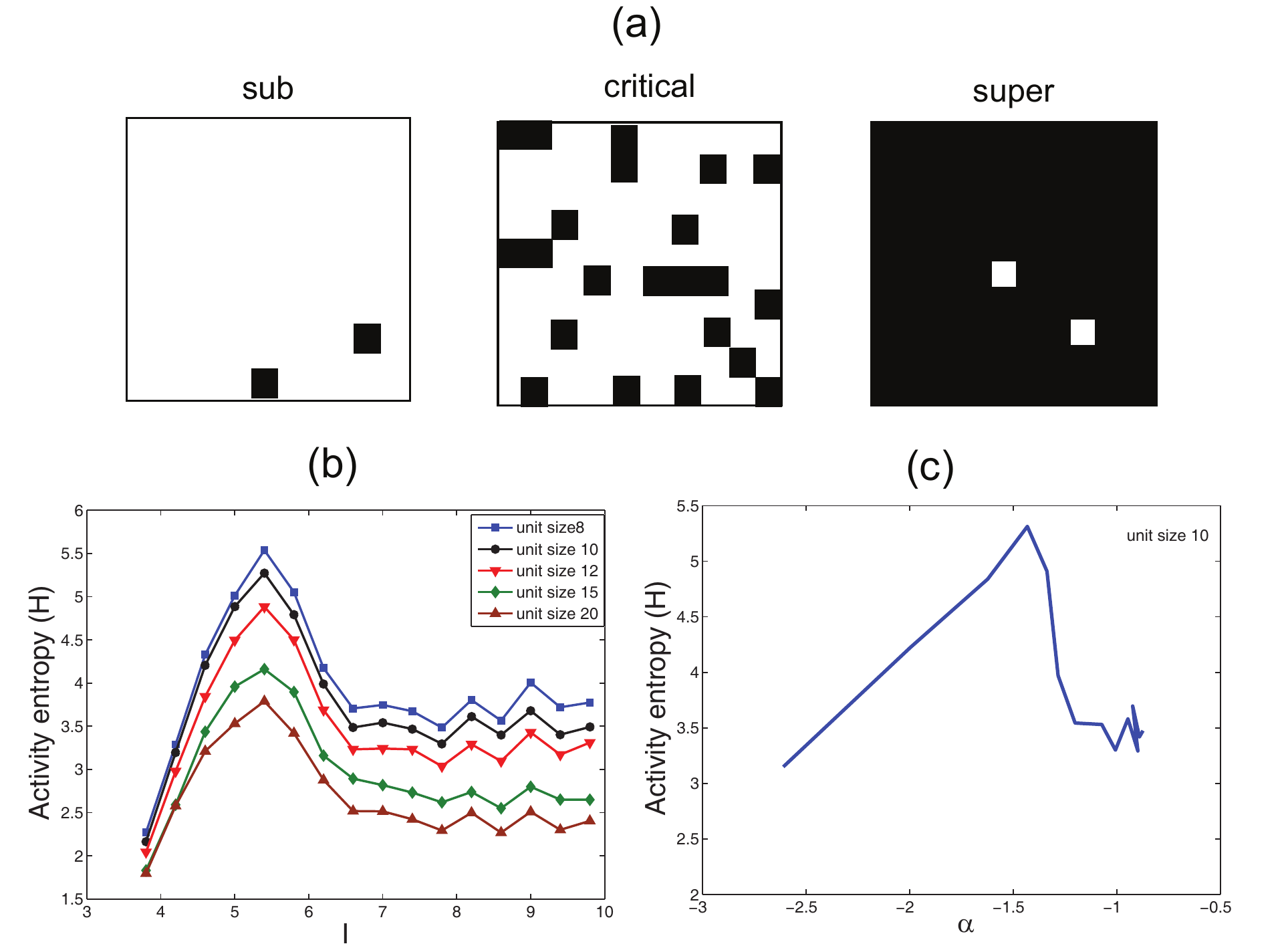}
 \caption{(a) Examples of activity binary patterns for subcritical, critical and supercritical states when the $100$ neurons are arranged as a $10\times10$ array. Active sites are marked as black squares. (b) The entropy of activity patterns is maximized when the stimulus current is about $5.5$, which is robust for the changes of unit size. Here time bin is set to be $4$~ms. (c) Entropy of activity patterns is maximized when $\alpha$ is $-1.5$ (i.e. critical state). Here unit size is $10$.}\label{fig4}
 \end{center}
\end{figure}

\begin{figure}
 \begin{center}
 \includegraphics[height=0.8\textwidth]{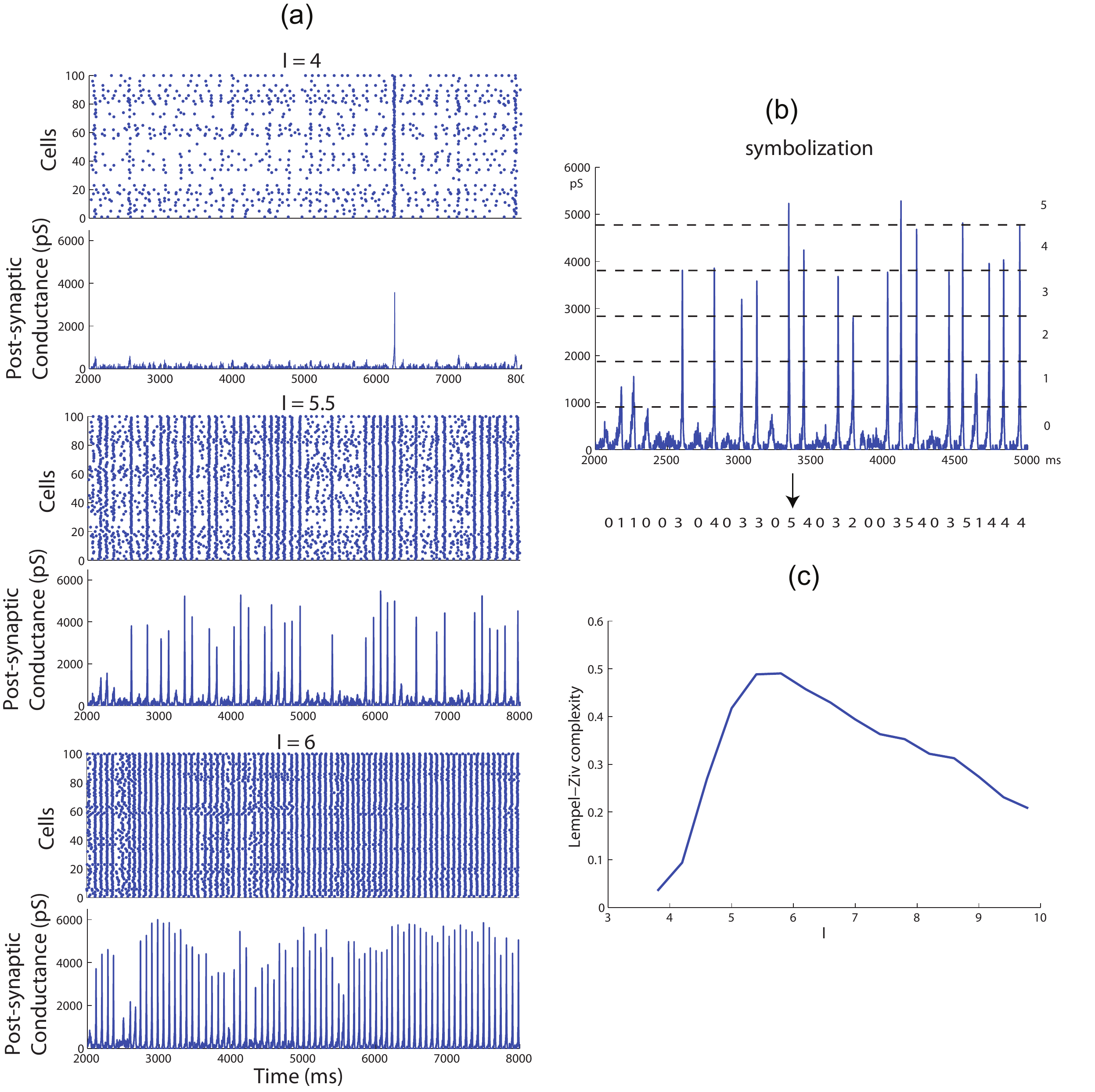}
 \caption{Complexity of synaptic conductance generated from the integration of network activity reaches maximum during the critical state. (a) Raters of network activity and corresponding post-synaptic conductance for different states. (b) Symbolization strategy of the synaptic conductance signal. (c) Lempel-Ziv complexity of the symbolized synaptic data is maximal at the critical state ($I=5.5$).}\label{fig5}
 \end{center}
\end{figure}

Examples of population events represented as binary patterns are shown in Fig.~\ref{fig4} (a), which indicate that temporally sub-synchronous neuronal activity occurred during the critical state can induce complex distributions of spacial patterns. In order to measure the complexity of activity patterns, entropy $H$ is introduced and defined as $H=-\sum_{i=1}^n p_{i}\log_{2}p_{i}$, where $n$ is the number of unique binary patterns and $p_{i}$ is the probability that pattern $i$ occurs \cite{shew2011information}. For calculation convenience neuronal activities are measured in pattern units consisting of a certain number of neurons. In each time bin if any cell of the unit is firing, the event of this unit is active; otherwise it is inactive. Fig.~\ref{fig4} (b) shows that the activity entropy of neuronal population are obviously maximized when the external current $I$ is about $5.5$ where neuronal avalanches occur ($\alpha=-1.5$ for Fig.~\ref{fig4} (c)). This result is robust for the changes of unit size and highly consistent with the analysis of cortical in-vitro recordings studied in \cite{shew2011information}. To investigate how the complex population events of this neural network influence downstream neurons, we calculate the summation of all post-synaptic conductance integrated from each pre-synaptic neuron in the network (Fig.~\ref{fig5} (a)). Each single synapse is the standard kinetic model as described in \cite{destexhe1998kinetic} with the peak conductance of $0.1$~nS, the rise and decay time constant of $0.5$~ms and $2$~ms respectively. Then time series of synaptic conductances are symbolized as numbers between $0\sim5$ by equally sorting the peak values of synaptic signals into six levels (Fig.~\ref{fig5} (b)). Based on the symbolized data we calculate the Lempel-Ziv complexity for cases with various external current (shown in Fig.~\ref{fig5} (c)). Lempel-Ziv complexity is used to quantify distinct patterns in symbolic sequences especially binary signals \cite{lempel1976complexity}. The peaked Lempel-Ziv complexity at about $I=5.5$ where critical dynamics exist further illustrates that information can be maximally transmitted in neural networks in forms of neuronal avalanches. The sub- and supercritical events of neuronal population produce either weak or regular post-synaptic signal with less encoded information that can be propagated to downstream neurons.

\begin{figure}
 \begin{center}
 \includegraphics[height=0.8\textwidth]{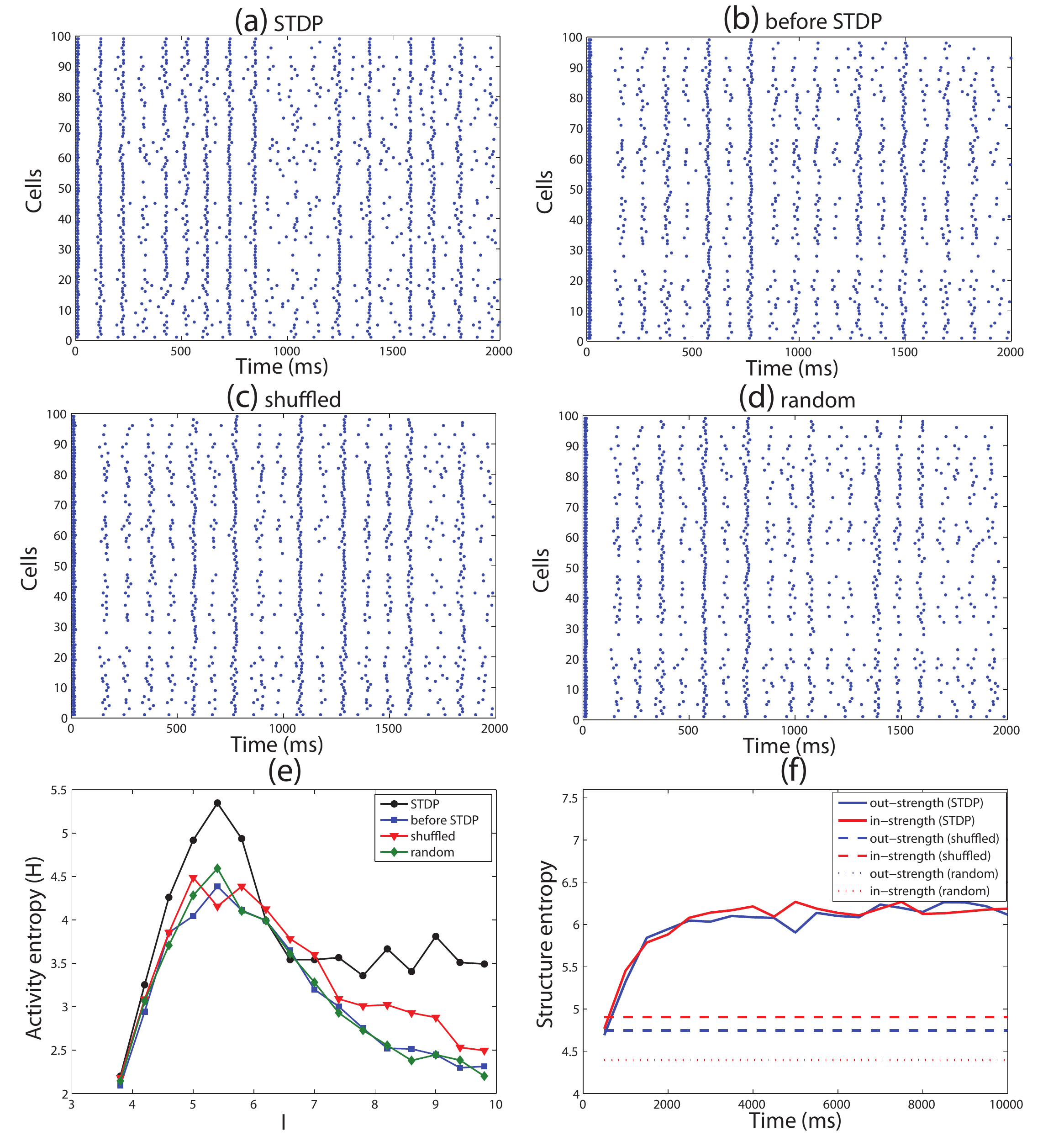}
 \caption{Entropy of network activity and structure for networks with different topologies. (a) Rasters of network activity for different networks (where $I=5.5$): (a) STDP network; (b) network before STDP update; (c) network with shuffled synaptic weights of the STDP network; (d) network with randomly distributed synaptic weights. All of these networks share the same individual neurons. (e) The STDP network exhibits much larger activity entropy than the other networks. The unit size is $10$.(f) Structure entropy of both inward and outward synaptic weights are greatly increased through the STDP update.}\label{fig6}
 \end{center}
\end{figure}

\begin{figure}
 \begin{center}
 \includegraphics[height=0.25\textwidth]{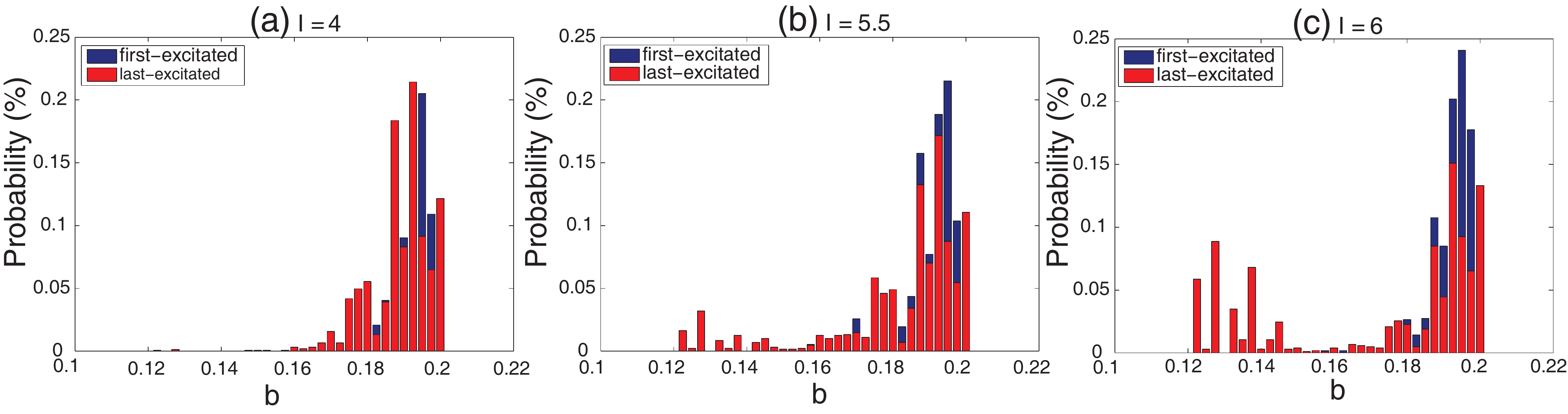}
 \caption{Structure properties of the STDP network. (a)$\sim$(c) Relationship between the individual neuron's excitability (represented by the parameter $b$) and its probability of being first excited (blue) or last excited (red) for different states.}\label{fig7}
 \end{center}
\end{figure}

\begin{figure}
 \begin{center}
 \includegraphics[height=0.5\textwidth]{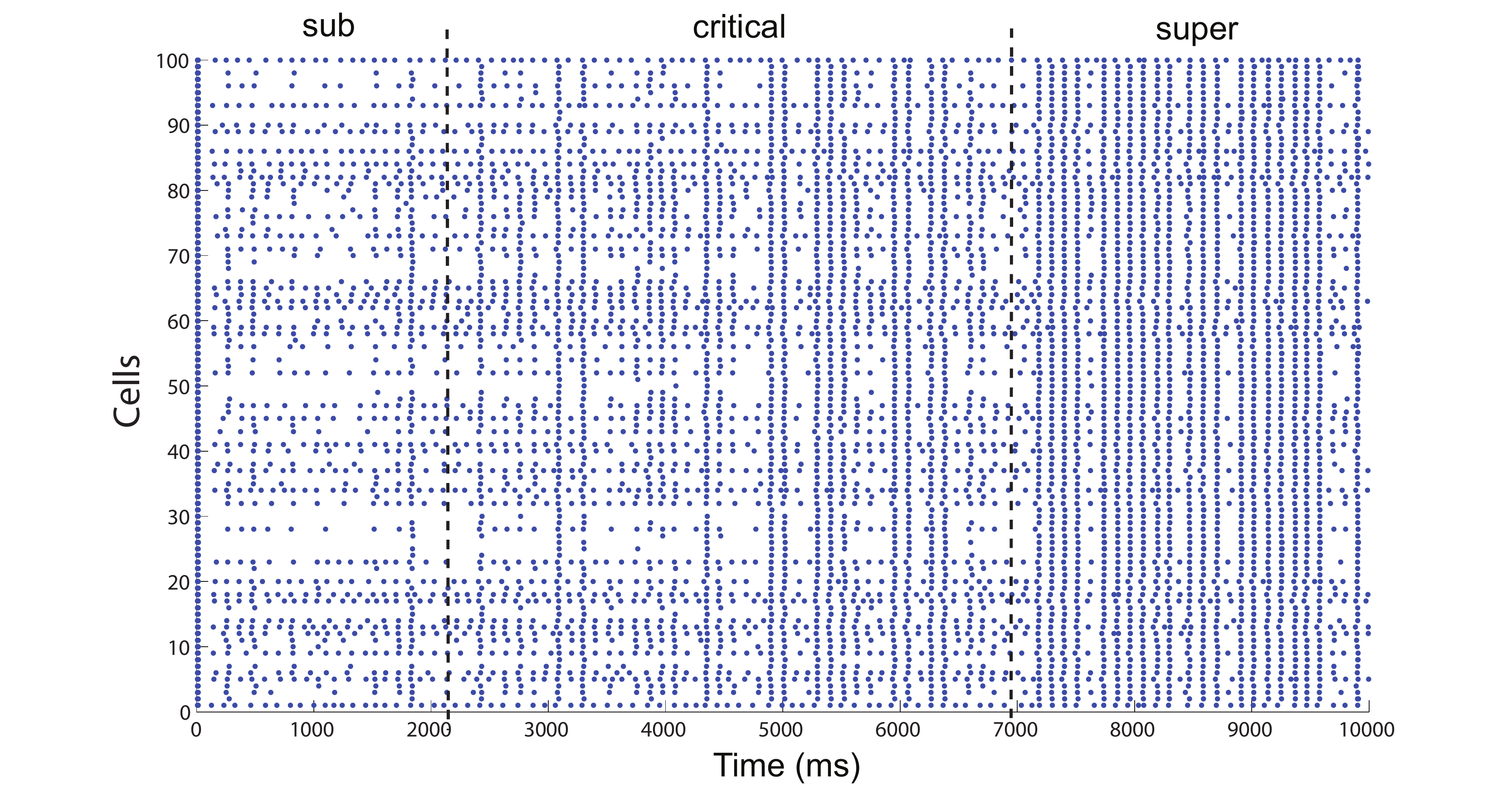}
 \caption{Raster of neuronal activity during STDP learning process. Network dynamics indicate that the dynamic of neuronal activity has been changed from the subcritical state to the critical state, and finally to the supercritical state during the update of network structure. Here the driving current is $I=6$.}\label{fig8}
 \end{center}
\end{figure}

To examine the influence of self-organized network connectivity updated by synaptic plasticity (STDP) on the network population events, we compare the critical dynamics of neuronal activities for four different networks (Fig.~\ref{fig6} (a)$\sim$(d)). STDP network is the network updated by the STDP learning rule as described previously in this paper; before STDP update the network has globally coupled network with constant synapses $g_{max}/2$, where $g_{max}$ is the maximal value of synaptic conductance used in the STDP network; random network has synapses uniformly distributed in $[0,g_{max}]$; shuffled network has the same synaptic distribution as STDP network but synaptic weights across neurons are shuffled. All these four types of network are composed of the same heterogeneous cells and share the same average synaptic weight being about $g_{max}/2$. Fig.~\ref{fig6} (e) shows that all of these networks reach the maximal activity entropy as neuronal avalanches occur when external current is about $5.5$. However, the activity entropy of the STDP network is much larger than that of the other networks, indicating the high information capacity of the STDP-refined neural network. Meanwhile, structure entropy is used to characterize the complexity of synaptic connectivity distribution, where the entropy has the same definition as the activity entropy but with $p_{k}$ as the probability that average value of inward or outward synaptic strength $g_{i}$ (normalized) lies within bin $k$ ($k\in[0,1]$ with a step of $0.005$). It can be clearly seen from Fig.~\ref{fig6} (f) that structure entropy for both in-strength and out-strength of the STDP network is dramatically enhanced during the synaptic refinement process. The reason is, the distinct differences of both inward and outward links between active neurons and inactive neurons in the STDP network make the mean inward or outward synaptic strengths still broadly distributed within the range of $[0,1]$ even after averaging, whereas for the other networks the mean inward or outward synaptic strengths are averaged to be around $0.5$ with little diversity. The active-neuron-dominant structure with high complexity actually encodes the intrinsic heterogeneity of individual neuron's dynamics. This complex network structure in turn contributes to the high activity entropy of neuronal population as shown in Fig.~\ref{fig6} (e). Fig.~\ref{fig7} (a)$\sim$(c) demonstrate that when external stimulation is strong enough to make the network reach the critical state, the active-neuron-dominant structure entails excitation to be firstly expressed on the most active neurons and then transferred to neighboring cells through the strong outward synaptic connections. Moreover, during the STDP synaptic updating process the dynamic of neuronal activity has been changed from the subcritical state to the critical state, and finally to the supercritical state (Fig.~\ref{fig8}). This indicates that neuronal avalanches may be natural phenomenon emergent from synaptic organization of neural networks.

\section{conclusion}

In this paper, information transmission of a self-organized neural network with active-neuron-dominant structure is investigated. Neuronal avalanches where activity propagation obeys a power-law distribution of population event sizes with an exponent of $-3/2$ can be observed in this network. During this critical dynamic state, excitation of network activity originating from the most active neurons can be spread efficiently and exhibits the largest activity entropy, thus providing the most complex post-synaptic inputs to downstream neurons. Specifically, the active-neuron-dominant structure emerging from synaptic plasticity dramatically increases both the activity and the structure entropy compared with alternative archetypal networks with different neural connectivity. This self-refinement of network structure is beneficial for improving information capacity of neural network.

A recent study reports that information retention in neuronal avalanches can be optimized by appropriately skewing synaptic weight distribution, with many weak weights and only a few strong ones \cite{chen2010few}. This kind of network model correctly matches the activity patterns observed in experimental data. Similarly, in our model the STDP learning process selectively strengthens the outward connections of active neurons with high excitability and meanwhile weakens or removes their redundant inward connections. The resulted distribution of refined synaptic weights is highly consistent with the above mentioned observations in \cite{chen2010few} and further demonstrates its effect on improving information capacity of neural circuits. Besides the measure of entropy, another efficient measure of the complex dynamics accompanied adaptive behavior called Causal Connectivity Analysis was proposed and used to predict the functional consequences of neural network lesions \cite{seth2005causal}. They suggested that lesions to input neurons with strong causal projections to outputs have a profound impact on network function. Moreover, learning and memory has been observed during the critical state in a computational study \cite{de2010learning}. Neuronal activity as a collective process requires flexible and efficient branched paths of signal propagation, which could be achieved by self-organization of neural network. The effect of synaptic adaptation involving both facilitation and depression in short-term plasticity on the occurrence of neuronal avalanches could be investigated in our future work.

\begin{acknowledgments}
We thank Dr. X. Xu for his valuable discussion. This research was funded by a Hong Kong University Grants Council Grant Competitive Earmarked Research Grant (CERG) number PolyU $5279/08$E.
\end{acknowledgments}



\end{document}